\documentclass[12pt]{article}
\usepackage{amsmath}
\usepackage{graphicx,psfrag,epsf}
\usepackage{enumerate}
\usepackage{natbib}
\usepackage{rotating}
\usepackage{amstext}
\usepackage{array} 
\usepackage{longtable}
\usepackage{epsfig} 
\usepackage{epstopdf} 
\usepackage{amssymb}
\usepackage{array}
\usepackage{multirow}
\usepackage{booktabs}
\usepackage{mathtools}
\usepackage[hang,small,bf]{caption}
\usepackage{setspace}
\usepackage{natbib}
\usepackage{url}
\usepackage{hyperref}
\usepackage{bbm}
\usepackage{listings}
\usepackage{adjustbox}
\usepackage{chngcntr}
\usepackage{mathrsfs}
\usepackage{bbm}
\usepackage{placeins}
\usepackage[T1]{fontenc}
\usepackage{authblk}
\numberwithin{equation}{section}
\counterwithin{table}{section}
\counterwithin{figure}{section}
\usepackage{color}
\usepackage{subcaption}
\usepackage[usenames,dvipsnames,svgnames,table]{xcolor}
\usepackage{hyperref}
\hypersetup{
     colorlinks   = true,
     linkcolor    = red,
     citecolor    = blue
}

\newcommand{\blind}{0}

\begin{document}

\def\spacingset#1{\renewcommand{\baselinestretch}%
{#1}\small\normalsize} \spacingset{1}


\if0\blind
{
  \title{\bf Bayesian inference for generalized extreme value distribution 
  				with Gaussian copula dependence}
  \author{Bo Ning\hspace{.2cm}\\
    Department of Statistics, North Carolina State University\\
    and \\
    Peter Bloomfield \\
    Department of Statistics, North Carolina State University}
  \maketitle
} \fi

\if1\blind
{
  \bigskip
  \bigskip
  \bigskip
  \begin{center}
    {\LARGE\bf Title}
\end{center}
  \medskip
} \fi

\bigskip
\begin{abstract}

Dependent generalized extreme value (dGEV) models have attracted much
attention due to the dependency structure that often appears in real
datasets.
To construct a dGEV model, a natural approach is to assume that some parameters in the
model are time-varying.
A previous study has shown that a dependent Gumbel process can be naturally
incorporated into a GEV model.
The model is a nonlinear state space model with
a hidden state that follows a Markov process,
with its innovation following a Gumbel distribution.
Inference may be made for the model using Bayesian methods,
sampling the hidden process
from a mixture normal distribution, used to approximate the Gumbel distribution.
Thus the response follows an approximate GEV model.
We propose a new model in which each marginal distribution is an
exact GEV distribution.
We use a variable transformation to combine the marginal CDF of a Gumbel
distribution with the standard normal copula.
Then our model is a nonlinear state space model in which the hidden state equation
is Gaussian.
We analyze this model using Bayesian methods,
and sample the elements of the state vector using
particle Gibbs with ancestor sampling (PGAS).
The PGAS algorithm turns out to be very efficient in 
solving nonlinear state space models.
We also show our model is flexible enough to incorporate seasonality. 
\end{abstract}

\noindent%

\spacingset{1.45}

\section{Introduction}
\label{gev-int}

Generalized extremely value models have been used extensively to estimate and 
predict extreme events in many applications,
for instance in the environment, engineering, economics and finance. 
The theoretical properties of this model 
have been studied in the literature \citep{pick75, cast04}.
Both frequentist and Bayesian approaches have been developed to 
estimate those models \citep{cole01, cole96b}.
Recently, dependent GEV models (for short, dGEV models) 
have attracted much attention.
Those models are useful in applications when
extreme events are time-correlated.
For example, higher temperature in one day may be followed by 
a higher temperature in another day.

A GEV distribution is shown as follows:
\begin{equation}
\label{equ-1.1}
    P(Y \le y)
    =
    F(y) 
    = 
    \exp \left\{ -\left(1+\xi \frac{y - \mu}{\psi} \right)_+^{-\frac{1}{\xi}} \right\},
\end{equation}
where $\mu$, $\psi$ and $\xi$ are the location, scale and shape parameters respectively.
To extend the GEV model to a dGEV model,
it is common to introduce dependence for some of its parameters.
Previous studies have focused on
letting $\mu$ follow an AR(1) process with other
parameters time invariant \citep{huer07};
and letting $\mu$, $\psi$ and $\xi$
be time-dependent and follow ARMA-type process \citep{wei16}.
However, since we consider the dependency of extreme events,
a more general way to introduce the time-dependent structure
is to insert a copula;
this idea is originally proposed by \citet{naka11, naka15}.
To be explicit,
their model is constructed as follows.

If we define
\begin{equation}
\label{equ-1.2}
  \alpha_t 
  \equiv 
  \log \left\{\left(1 + \xi \frac{Y_t - \mu}{\psi}\right) _+^ \frac{1}{\xi} \right\},
\end{equation}
then $\alpha_t$ follows the Gumbel distribution.
Solving (\ref{equ-1.2}) for $Y_t$,
we find the following expression,
\begin{equation}
\label{equ-1.3}
  Y_t 
  = 
  \mu + \psi \frac{\exp(\xi \alpha_t) - 1}{\xi},
\end{equation}
and if $\alpha_t$ is a hidden AR(1) process,
\begin{equation}
\label{equ-1.4}
  \alpha_{t+1} = \phi \alpha_t + \eta_t, \quad \eta_t \sim \text{Gumbel},
\end{equation}
then the model becomes a nonlinear non-Gaussian state space model.
To estimate the parameters in the model, 
they first show that the model can be approximated by
a linear Gaussian state space model,
by finding a linear approximation for nonlinear observation equations
and using mixed normal distributions to approximate the Gumbel errors
in the state equation. 
However, 
from (\ref{equ-1.4}) the marginal distribution of $\alpha_t$ does not follow a
Gumbel distribution,
because the sum of two Gumbel-distributed random variables in general
does not follow a Gumbel distribution.
We propose a method to allow $\alpha_t$ to follow exactly the
Gumbel distribution.
The model has Gumbel marginal distributions linked together with a
Gaussian copula with $\text{AR}(1)$ dependence.
By this construction, 
the marginal distribution of $\alpha_t$ is exactly from 
a Gumbel distribution. 
The details are shown in Section~\ref{gev-mod}.

Because the number of variables in the dGEV model is large,
we use Bayesian analysis to sample draws from the joint distributions.
Markov chain Monte Carlo (MCMC) allows us to sample from the conditional
distribution for each individual parameter,
and thus we only need to take care of the conditional distribution of
each parameter at one time instead of the joint distribution for all parameters.
A drawback of using the MCMC algorithm is that
the mixing can be slow due to the high dimensional 
state vector in the model.
As a result, many iterations are needed
and methods to reduce the correlation between chains such as
marginalization, permutation and
trimming \citep{dyk08, liu94, liu94b}
must be considered.

In Bayesian analysis, a common technique is to use 
Kalman filtering and backward smoothing (KFBS)
\citep{west97} for sampling draws from 
linear Gaussian state space models. 
For nonlinear non-Gaussian state space models, 
the KFBS cannot applied directly.
One way of sampling those model is to
find a linear approximation for the
nonlinear equations \citep{shep97};
common techniques include 
using a first-order Taylor series expansion based on the observation
equations,
or a second-order Taylor expansion based on the log-likelihood functions,
and finding a mixed normal approximation for the non-Gaussian errors.
The sampling approach is to apply KFBS algorithm on the approximated
linear Gaussian state space model,
and using Metropolis-Hastings (M-H) to accept or reject the draws.
However, 
When the model is highly nonlinear and non-Gaussian 
these approximation techniques may perform very poorly;
as a result, the M-H step will reject most of the draws,
thus increasing the computation time and reducing the mixing rate of the chain.
There are many extensions for the KFBS for adapting to a nonlinear state space model
including extended Kalman filter, robust extended Kalman filter \citep{eini99}
and using mixed normal densities to approximate nonGaussian densities
\citep{niem10}.

An alternative way to make inferences in a nonlinear state space model
is by using 
sequential Monte Carlo (SMC).
This approach is as follows: 
it first generates particles from a chosen proposal density;
and then calculate the weights
between the proposal density and the true density. 
The generated particles and their weights give a discrete distribution approximation
to the true density.
However, one issue with this method is the degeneration of the weights,
thus a resampling step is needed when calculating the weights.
This method is also known as importance sampling.
Recently, much effort has been put into incorporating SMC within MCMC;
some of the methods include particle Metropolis-Hastings and the
particle Gibbs sampler \citep{andr10}.
To use MCMC approach is more convenient when dealing with a model with 
many variables.
The particle Gibbs sampler is the method we shall use in this paper;
this method uses collapsed Gibbs sample \citep{liu94b}, 
which allows the joint distribution to be invariant in MCMC sampling.
The essential idea is to combine a conditional SMC (cSMC) 
\citep{liu01} with the Gibbs sampler.
The cSMC method needs to keep a path
known as ancestor lineage to survive in each resampling step. 
However, this algorithm has two weaknesses:
poor mixing rate and particle degeneration \citep{douc11, chop15}.
An extension to address these weaknesses is to do backward smoothing,
which known as forward sampling and 
backward smoothing (FFBS) \citep{whit10, lind13, kant15}.
A drawback for the FFBS algorithm is that adding
the backward step increases the computation time.
A more recent method which is known as 
Particle Gibbs with ancestor sampling (PGAS)
\citep{lind14} is proposed to address the computation issues of FFBS.
The PGAS approach is as follows:
for each forward sampling step, 
instead of fixing the ancestor lineage,
the algorithm resamples the ancestor lineage.
In this case, some particles with small weights in the ancestor lineage
will drop out.

In the rest of the paper,
we will introduce our novel dGEV model,
which incorporates the dependence between extreme events
through a dependent Gumbel distribution;
the dependent Gumbel distribution is constructed with a
normal copula,
and in the final model the observed variable is sampled from an exact GEV distribution.
Our model can be expressed as a nonlinear Gaussian state space model;
we use PGAS to sample the hidden process in the nonlinear state space model.
We also show that our model may be extended to incorporate seasonal
components,
which are needed for some datasets;
we name the result a seasonal dGEV model. 

In the simulation study, we show our estimated posterior medians
for parameters are close to the true values, 
and the true values are contained in the 95\% credible intervals.
Because the latent variables marginally follow a standard
normal distribution,
we could use standard normal distribution tails to characterize the extremeness 
of the values. 

We fit the model to two real datasets: one is an annual maximum water
flow dataset and the other is a weekly minimum return for the S\&P 500
index.

The remaining sections are arranged as follows:
Section~\ref{gev-mod} introduces our dGEV model;
Section~\ref{gev-pos} describes a Bayesian computation steps;
Section~\ref{gev-tvg} adds seasonality on the dGEV model;
Section~\ref{gev-sim} illustrates the dGEV and seasonal dGEV models using simulated data;
Section~\ref{gev-rea} conducts a real data study for a water flow dataset and a financial dataset;
Section~\ref{gev-con} concludes.

\section{The dGEV model.}
\label{gev-mod}

Our dGEV model is constructed as follows:
suppose that $\beta_t$ follows the standard normal distribution, so that
$\Phi(\beta_t)$ follows the uniform $(0, 1)$ distribution.
Let $G(\alpha)$ be the CDF of the Gumbel distribution:
$G(\alpha) = \exp( - \exp( - \alpha))$.
If
\[
    \alpha_t = G^{-1}(\Phi(\beta_t))
    =
    - \log( -\log(\beta_t)),
\]
then $\alpha_t$ follows the Gumbel distribution.

Now suppose that $\beta_t$ follows an AR(1) process:
\begin{equation}
\label{equ-beta-ar1}
   \beta_{t+1} 
     =
      \phi \beta_t 
      + 
      \eta_t, \quad \text{ where } \eta_t \sim N(0, 1-\phi^2)
\end{equation}
for $t = 1, \dots, T-1$ and $\beta_1 \sim N(0, 1)$,
and let
\begin{eqnarray}
    \notag
    Y_t 
    &  =  &
      \mu 
      + 
      \psi
      \frac{
          \exp (\xi \alpha_t) - 1
      }{
          \xi
      }
      + \epsilon_t
    \\
\label{equ-observation}
    &  =  &
      \mu 
      + 
      \psi \frac{(-\log(\Phi(\beta_t)))^{-\xi} - 1}{\xi}
      + \epsilon_t,
      \quad \epsilon_t \sim N(0, \sigma^2)
\end{eqnarray}
where $t = 1, \dots, T$ and $\beta_1 \sim N(0, 1)$.
The normal error $\epsilon_t$ in the observation equation is a added term
that makes the model have a nonlinear state space presentation;
the error should be have a small variance to make it negligible.

The parameters in our model are
$\mu$, $\psi$, $\xi$, $\theta$, $\sigma^2$ and the latent variables $\beta_{1:T}$.
In Bayesian analysis,
we adopt a similar priors for these parameters to those suggested by
\citep{cole96, chav12}:
$\mu \sim N(\nu_\mu, \sigma^2_\mu)$, 
$\psi \sim \text{Gamma}(a_\psi, b_\psi)$, 
$\xi \sim N(\nu_\xi, \sigma^2_\xi)$.
The prior for $\phi \sim U(-1, 1)$ to make sure the AR(1) process in (\ref{equ-beta-ar1})
is stationary.
Last, the prior for $\sigma^{-2} \sim \text{InvGamma}(a, b)$,
which we later abbreviate $\text{IG}(a, b)$.

\section{Posterior computation}
\label{gev-pos}

In this section,
we describe a MCMC algorithm for sampling draws of the parameters 
in the dGEV model. 
Because of the many parameters,
a slow mixing rate will cause the chain converge very slowly.
To ameliorate this problem,
we consider using trimming and blocking techniques;
\citet{dyk08} have justified those techniques
as useful.

In the following, 
to simplify the notation, 
let $\boldsymbol \theta$ to be any parameter(s) in the model and 
$-\boldsymbol \theta$ to be the parameters in the model except 
$\boldsymbol \theta$.
The outline of the MCMC algorithm is designed as follows:
we sample 
$\pi_{-\boldsymbol \vartheta}(\boldsymbol \vartheta| \boldsymbol Y)$
from its posterior distribution by grouping $\mu$, $\psi$, and $\xi$ together and treating
them as a block;
and then sample $\pi_ {-\sigma^2}(\sigma^2| \boldsymbol Y)$,
$\pi_{-\phi}(\phi| \boldsymbol Y)$ from the conditional posterior distribution 
using Gibbs and Metropolis-Hastings algorithm;
last, 
sample $\pi_{-\boldsymbol \beta}(\boldsymbol \beta| \boldsymbol Y)$ using 
particle Gibbs with ancestor sampler (PGAS), which will be introduced later.

\subsection{Sampling $\mu$, $\psi$, $\xi$}

It is suggested to group $\mu$, $\psi$, $\xi$ parameters together 
and sample from their joint distribution to reduce 
the correlation between MCMC chains \citep{naka11}.
Then the joint conditional posterior distribution for  
$\boldsymbol \vartheta$ is
\begin{equation}
\label{equ-post-mu-psi-xi}
	\pi_{-\boldsymbol \vartheta}(\boldsymbol \vartheta|\boldsymbol Y)
	=
	\prod_{t=1}^T 
	\pi_{-\boldsymbol \vartheta}(\boldsymbol \vartheta|Y_t)
	\pi(\mu) \pi(\psi) \pi(\xi)
\end{equation}

Although the priors for $\boldsymbol \mu$ and 
$\boldsymbol \psi$ is conjugate with their likelihoods, 
the prior for $\boldsymbol \xi$ does not.
When we group them as a trivariate parameters, 
the joint likelihood is not conjugacy with the priors,
thus we use Metropolis-Hastings algorithm 
and find a proposal density using the second-order Taylor expansion of
$\log \pi_{-\boldsymbol \vartheta}(\boldsymbol \vartheta| \boldsymbol Y)$
as follows.
Let $\boldsymbol \vartheta^* \sim \text{MVN}(\boldsymbol \nu^*, \boldsymbol \Sigma^*)$,
where 
\begin{eqnarray}
\notag
	& \boldsymbol \Sigma^{*-1}&
	= 
	- \frac{\partial^2 
	\log \pi_{-\boldsymbol \vartheta}(\boldsymbol \vartheta| \boldsymbol Y)}
	{\partial \boldsymbol \vartheta \ \partial \boldsymbol \vartheta^T} 
	\Big |_{\boldsymbol \vartheta = \hat{\boldsymbol \vartheta}},\\
\notag
	& \boldsymbol \nu^* &
	= 
	\hat{\boldsymbol \vartheta} 
	+ 
	\boldsymbol \Sigma^*
	\cdot
	\frac{\partial 
	\log \pi_{-\boldsymbol \vartheta}(\boldsymbol \vartheta| \boldsymbol Y)}
	{\partial \boldsymbol \vartheta} 
						\Big |_{\boldsymbol \vartheta = \hat{\boldsymbol\vartheta}},
\end{eqnarray}
and $\hat{\boldsymbol \vartheta}$ is a value that maximizes the posterior of 
$\pi_{-\boldsymbol \vartheta}(\boldsymbol \vartheta| \boldsymbol Y)$;
this value can be calculated by using \textsf{optim} in \textsf{R}.
After sampling a draw
$\boldsymbol \vartheta^*$
from this proposal density,
we use Metropolis-Hastings to accept or reject
$\boldsymbol \vartheta^*$
with ratio
\begin{equation}
\notag
	\alpha(\boldsymbol \vartheta, \boldsymbol \vartheta^*) 
	= 
	\min \Big\{1, 
	\frac{\pi_{-\boldsymbol \vartheta}(\boldsymbol \vartheta^*| \boldsymbol Y)
			f(\boldsymbol \vartheta| \boldsymbol \nu^*, \boldsymbol \Sigma^{*})}
	{\pi_{-\boldsymbol \vartheta}(\boldsymbol \vartheta| \boldsymbol Y)
			f({\boldsymbol \vartheta^*| \boldsymbol \nu^*, \boldsymbol \Sigma^{*}})}    
	\Big\}.
\end{equation}

\subsection{Sampling $\sigma^2$}

To sample $\sigma^2$,
we choose a conjugate prior inverse gamma prior as follows,
\begin{equation*}
	\sigma^{2} \sim \text{IG}(a, b).
\end{equation*}
Since $\epsilon_t$ should be small, 
we choose $a$ to be large and $b$ to be small so that the prior will have a smaller
variance.

The posterior of $\sigma^{2}$ can be written as follows:
\begin{eqnarray}
\label{equ-post-sigsq}
	\pi_{-\sigma^2}(\sigma^2|\boldsymbol Y)
	& = &
	\text{IG} \left(
        a + \frac{T}{2}, 
        b +
	\frac{\sum_{t=1}^T \left( y_t - \mu - \frac{\psi}{\xi}
        (-\log(\Phi(\beta_t))^{-\xi} - 1) \right)^2}{2} 
        \right).
\end{eqnarray}

\subsection{Sampling $\phi$ }

The conditional posterior of $\phi$ can be expressed as follows:
\begin{equation}
\label{equ-post-phi}
   \pi_{-\phi}(\phi| \boldsymbol Y)  = (2\pi)^{-T/2}(1-\phi^2)^{-T/2} \cdot
      \exp\Big\{ 
               -  \frac{\sum_{t=2}^T (\beta_t - \phi \beta_{t-1})^2}{2(1 - \phi^2)} 
             \Big\} \cdot \pi(\phi)
\end{equation}
where $\pi(\phi)$ is the prior of $\phi$. 
Because the posterior density is not a well known distribution,
similar to dealing with parameter $\boldsymbol \vartheta$, 
we find a proposal density for $\pi_{-\phi}(\phi| \boldsymbol Y)$ as
$q_{-\phi}(\phi| \boldsymbol Y) \sim N(\nu_\phi, \sigma_\phi^2)$ 
with
\begin{eqnarray}
\notag
	& \sigma_\phi^{-2}&
	= 
	- \frac{\partial^2 \log \pi_{-\phi}(\phi| \boldsymbol Y)}
	{\partial \phi^2} 
	\Big |_{\phi = \hat{\phi}},\\
\notag
	& \nu_\phi &
	= 
	\hat{\phi} 
	+ 
	\sigma_\phi^2
	\cdot
	\frac{\partial 
	\log \pi_{-\phi}(\phi| \boldsymbol Y)}
	{\partial \phi} 
						\Big |_{\phi = \hat{\phi}}.
\end{eqnarray}

After we draw 
$\phi^* \sim q_{-\phi}(\phi| \boldsymbol Y) \sim N(\nu_\phi, \sigma_\phi^2)$ ,
$\phi^*$ is accepted with probability
\begin{equation}
\notag
	\alpha(\phi, \phi^* ) 
	= 
	\min \Big\{1, 
	\frac{\pi_{-\phi}(\phi^*| \boldsymbol Y)
			q_{-\phi}(\phi| \boldsymbol Y)}
	{\pi_{-\phi}(\phi| \boldsymbol Y)
			q_{-\phi}(\phi^*| \boldsymbol Y)}    
	\Big\}.
\end{equation}

\subsection{Sampling $\beta_{1:T}$}

The posterior distribution for 
$\pi_{- \boldsymbol \beta}(\boldsymbol\beta| \boldsymbol Y)$ 
can be expressed as the following,
\begin{eqnarray}
\notag
	\pi_{- \boldsymbol \beta}(\boldsymbol\beta| \boldsymbol Y)
	= 
	\prod_{t=1}^T f(Y_t| \beta_t, \boldsymbol \vartheta, \sigma^2)
	\prod_{t=2}^T f(\beta_t| \beta_{t-1}, \phi)
	\pi(\beta_1),
\end{eqnarray}
with
\begin{eqnarray*}
	f(Y_t| \beta_t, \boldsymbol \vartheta, \sigma^2) 
	& = &
	\Big (\frac{1}{2\pi\sigma^2} \Big)^{-\frac{T}{2}}
	\exp \Big\{  -
		\frac{Y_t - \mu -  \frac{\psi}{\xi} ((-\log(\Phi(\beta_t)))^{-\xi}-1)}
		{2\sigma^2}
	\Big \},\\
	f(\beta_t| \beta_{t-1}, \phi) 
	& = &
	\Big (\frac{1}{2 \pi(1-\phi^2)} \Big)^{-\frac{T-1}{2}}
	\exp \Big\{ -
		\frac{\sum_{t=2}^T (\beta_t - \phi \beta_{t-1})^2}{2 (1-\phi^2)}
	\Big\},\\
	\pi(\beta_1) 
	& = &
	-\frac{1}{\sqrt{2 \pi}} 
	\exp\Big\{  
		- \frac{\beta_1^2}{2}
	\Big\}.
\end{eqnarray*}

Due to the highly nonlinear expression of the likelihood 
$f(Y_t| \beta_t, \boldsymbol \vartheta, \sigma^2)$,
the Kalman filter and backward smoothing algorithm 
developed based on linear state space models cannot be applied directly. 
That is because the linear approximation using Taylor expansion has very 
poor performance. 
As a result, the acceptance probability is almost $0$ for our model.

The particle Gibbs sampler proposed by \citet{andr10} provides an alternative 
to sampling the nonlinear state space model.
It is based on the conditional sequential Monte Carlo (cSMC) \citep{liu01} algorithm
which is similar to the SMC algorithm except that it assigns an
ancestor lineage keep out of the resample stage.
The particle Gibbs sampler treats the particles and weights generated from cSMC
as a discrete distribution approximating the true density;
a draw is randomly sampled from these particles according to the their weights.
Since the draws are obtained from an approximate distribution
and not the real one,
a pre-specified ancestor lineage or trajectory path 
is used to guide draws from the invariant unconditional distribution.
However, this method is known to have degenerate issues and a poor mixing rate.
A finer algorithm is to allow the pre-defined ancestor lineage to update during
the forward sampling process,
and drop some lineage with degenerate weights.
The method is known as particle Gibbs with ancestor sampler (PGAS), proposed 
by \citet{lind14}.
The method uses a trimming technique \citep{dyk08} 
to improving the mixing rate of a Gibbs sampler;
also, the degenerate weight issue is ameliorated.

In the remainder of this section, we shall introduce this method. 
We summarize the notations to be used later for convenience:
let $q_{- \boldsymbol \beta}(\beta_t| Y_t)$ 
be a proposal density,
$\beta_t^1, \dots, \beta_t^N$ be $N$ particles drawn from the proposal density,
$k$ is a index in $\{1, \dots, N\}$ and $\boldsymbol k = (1, \dots, N)$;
let $\boldsymbol \beta^k = (\beta_1^k, \dots, \beta_T^k)$ 
stand for the samples generated from the same trajectory path $k$;
let $\overline {\boldsymbol \beta}_t^k 
= 
\left( \beta_1^k, \dots, \beta_{t-1}^k, \beta_{t+1}^k, \dots, \beta_T^k
\right)$,
and 
$\overline {\boldsymbol k}
=
(1, \dots, k-1, k+1, \dots, N)$;
let $\beta_t'$ be the previous draws or the starting value for $\beta_t$
and $\beta_t^*$ be the current draws from 
$\beta^{\boldsymbol k}_t$;
let $w_t(\beta_t^k)$ be the unnormalized weight for the $k^\text{th}$ particle of 
$\beta_t$ sampled from $q(\beta_t| -\boldsymbol \beta, Y_t, \beta_{t-1})$, 
and $W_t(\beta_t^k)$ be its normalized weight. 

We consider two methods to choose a proposal density;
the first method is to use Taylor expansion on the mean of the observation equation.
For simplicity we write (\ref{equ-observation}) as the following:
\begin{eqnarray*}
	Y_t &=& f_{\boldsymbol \vartheta} (\beta_t) + \epsilon_t, \\
	\beta_t & = & g_\phi(\beta_{t-1}) + \eta_t,
\end{eqnarray*}
with 
$f_{\boldsymbol \vartheta} (\beta_t)  
= 
\mu + \psi  ((-\log(\Phi(\beta_t)))^{-\xi} - 1) / \xi$
and
$g_\phi(\beta_{t-1}) = \phi \beta_{t-1}$;
then we approximate $f_{\boldsymbol \vartheta} (\beta_t)$ with 
\begin{equation*}
	f_{\boldsymbol \vartheta} (\beta_t) 
	= 
	f_{\boldsymbol \vartheta} \left( g_\phi(\beta_{t-1}^{\boldsymbol k}) \right)
	+
	f'_{\boldsymbol \vartheta} \left(
        g_\phi(\beta_{t-1}^{\boldsymbol k}) \right) 
	\left( \beta_t - g_\phi(\beta_{t-1}^{\boldsymbol k}) \right) / c.
\end{equation*}
Because in a nonlinear equation,
the second term in the Taylor expansion can be relatively 
large, the constant $c$ is used to control the approximated density.
We adjust the value $c$ to keep the weights from degenerating.
Then the proposal density then can be transformed into a Gaussian distribution,
\begin{equation*}
	q_{-\boldsymbol \beta}(\beta_t| \boldsymbol Y)  
	=
	N(\nu_{\beta_t}, \sigma^2_{\beta_t}) 
\end{equation*}
with
\begin{eqnarray*}
	\sigma^{-2}_{\beta_t}
	& = &
	\frac{1}{(1-\theta^2)} 
	+ 
	f'_{\boldsymbol \vartheta} (g_\phi(\beta_{t-1}^{\boldsymbol k}))^2 / \sigma^2 \\
	\nu_{\beta_t} 
	& = &
	\frac{\sigma^2_{\beta_t}}{(1-\theta^2)} g_\phi(\beta_{t-1}^{\boldsymbol k})
	+
	\frac{f'_{\boldsymbol \vartheta} (g_\phi(\beta_{t-1}^{\boldsymbol k}))^2}{\sigma^2}
	\left( Y_t - f_{\boldsymbol \vartheta} (g_\phi(\beta_{t-1}^{\boldsymbol k})) 
	\times 
	g_\phi(\beta_{t-1}^{\boldsymbol k}) \right)
\end{eqnarray*}

The second method is to ignore the $\epsilon_t$ in (\ref{equ-observation}),
then we write $\beta_t$ as a function of $y_t$ as follows, 
\begin{equation*}
	\beta_t = \Phi^{-1} 
        \left( 
        exp \left(
        - \left( 1 + \xi \frac{Y_t - \mu}{\psi} \right)^{-\frac{1}{\xi}}_+
        \right)
        \right)
\end{equation*}
where $\Phi^{-1}$ denotes the inverse standard normal distribution
function.
Then we choose a proposal density to be the $t$-distribution with 
mean $\beta_t$ and a small value of degrees of freedom, namely 5. 
This method does not apply to general state space models,
but can perform very well in our model;
this is because the $\epsilon_t$ in our model is small. 

After specifying the proposal density, the PGAS algorithm is as follows,
\begin{itemize}
	\item At current iteration $i$, given $\boldsymbol \beta'$
	\item For $t = 1$
	\begin{itemize}
		\item Draw $\beta_{1}^{1:N-1} \sim q_{- \boldsymbol \beta}(\beta_1| Y_1)$
		\item Let $\beta_1^{N} = \beta'_1$
		\item Calculate weight for $l = 1, \dots, N$ as
					\begin{equation*}
					\begin{split}
						w_1^l 
						&:= 
						\frac{ 
						f(Y_1| \beta_1^{l}, \boldsymbol \vartheta, \sigma^2) \pi(\beta_1^{l}) 
						} 
						{
						 q_{-\boldsymbol \beta}(\beta_1^l| Y_1)  
						 }\\
						W_1^l
						& =
						\frac{w_1^l}{\sum_{l=1}^l w_1^l }
					\end{split}
					\end{equation*}
	\end{itemize}
	\item For $t = 2, \dots, T$
	\begin{itemize}
		\item Resample $\beta_{t-1}^k$ according to the weight $W^k$
				 for $k = 1, \dots, N-1$, denote as $\tilde \beta_{t-1}^{k}$.
		\item Sample $\tilde \beta_{t-1}^{N}$ from $\beta_{t-1}^{\boldsymbol k}$ with 
					the associated weights:
					\begin{equation*}
						\frac{
						w_t^l f(\beta'_t| \beta_{t-1}^l, \phi)
						}
						{
						\sum_{l=1}^N w_t^l f(\beta'_t| \beta_{t-1}^l, \phi)
						}
					\end{equation*}
		\item Draw $\beta_t^{1, \dots, N-1} \sim q_{-\boldsymbol \beta}(\beta_t| y_t)$
		\item Let $\beta_t^{iN} = \beta'_t$
		\item Let $\beta_{1:t}^{l} := (\tilde\beta_{1:t-1}^{l}, \beta_t^{l})$, 
					for $l = 1, \dots, N$
		\item Calculate weight for $l = 1, \dots, N$ as
					\begin{equation*}
					\begin{split}
						w_t^l 
						&= 
						\frac{ 
						f(Y_t| \beta_t^{l}, \boldsymbol \vartheta, \sigma^2) 
						p(\beta_t^{il}| \tilde\beta_{t-1}^{il}, \phi) 
						} 
						{
						 q_{-\boldsymbol \beta}(\beta_t^{il}| Y_t)  
						 }\\
						W_t^l
						& =
						\frac{w_t^l}{\sum_{l=1}^l w_t^l }
					\end{split}
					\end{equation*}
	\end{itemize}
	\item Draw k from $1, \dots, N$ with the corresponding probability $W_T^{1, \dots, N}$
	\item Let $\boldsymbol \beta = \boldsymbol \beta^{k}$
	\item Set $\boldsymbol \beta' = \boldsymbol \beta$, and go to the next iteration.
\end{itemize}
Then from the algorithm, we obtain draws for $\boldsymbol \beta$.

\subsection{Markov Chain Monte Carlo}
\label{sec-3.5}

In summary, the MCMC algorithm is conducted as follows:

\begin{enumerate}
\item initialize all the parameters values $\boldsymbol \vartheta$,
$\boldsymbol \beta$, $\phi$, $\sigma^2$;

\item sample $\boldsymbol \vartheta = (\mu, \psi, \xi)$ from (\ref{equ-post-mu-psi-xi}),
and using M-H algorithm;

\item sample $\sigma^2$ from (\ref{equ-post-sigsq});

\item sample $\phi$ from (\ref{equ-post-phi}), and using M-H algorithm;

\item sample $\boldsymbol \beta$ using PGAS;

\item repeat steps 2 --- 5 until sufficient draws have been obtained.
\end{enumerate}

\section{Seasonal dGEV model}
\label{gev-tvg}

Seasonality can be easily incorporated into our model 
by adding sinusoids to any of the location,
shape or scale parameters.
This method is convenient since it has flexibility in the choice of the
number of sinusoids and 
it does not greatly increase the difficulty of making inferences about the model.

We start by adding two sinusoids to the location parameter;
the model is shown as follows:
\begin{equation}
\begin{split}
\label{equ-seasonal-mean}
    Y_t 
    &  = 
      \mu + a_1 \cos(\omega t) + a_2 \sin(\omega t) \\
    & \quad + 
      \psi
     \frac{(-\log(\Phi(\beta_t)))^{-\xi} - 1}{\xi} + \epsilon_t, \ 
               \epsilon_t \sim N(0, \sigma^2),\\
   \beta_{t+1} 
     & = 
      \phi \beta_t 
      + 
      \eta_t, \ \eta_t \sim N(0, 1-\phi^2),
      \end{split}
\end{equation}
where 
$\omega = 2 \pi f$ is the angular frequency,
$f$ is the number of cycles that occur for each period of time.
For a given dataset, 
$f$ is typically known;
for example, it has value 1/365.25 for annual variability in a daily dataset, or
1/4 for a annual variability in a seasonal dataset.
In equation~(\ref{equ-seasonal-mean}),
$t = 1, \dots, T$,
$a_1$ and $a_2$ are the coefficients for the two components of the sinusoid,
and we let $\boldsymbol a = (a_1, a_2)$.

To estimate the parameters in~(\ref{equ-seasonal-mean}),
we use similar priors to those for the parameters described in Section \ref{gev-pos}.
For the added parameters in the seasonal component in this model,
we treat $\omega$ as a known parameter.
We observe that $\boldsymbol a$ has a multivariate normal likelihood in the 
equation, 
we choose the prior for each element $a_i$ to be 
$N(0, \sigma^2_{a_i})$.
Then the posterior for $\boldsymbol a$ is as follows,
\begin{equation}
	\pi_{-\boldsymbol a}(\boldsymbol a| \boldsymbol Y)
	\sim
	\text{MVN}(\boldsymbol \nu_{\boldsymbol a}, \boldsymbol \Sigma_{\boldsymbol a})
\end{equation}
with
\begin{eqnarray*}
	\boldsymbol \Sigma_{\boldsymbol a}^{-1}
	=
	\sigma^{-2}
	(\sum_{t=1}^T
	\boldsymbol p_t \boldsymbol p'_t 
	+ 
	\sigma^2 \boldsymbol \Omega^{-1}) , 
	\quad
	\boldsymbol \nu_{\boldsymbol a}
	= 
	\boldsymbol \Sigma_{\boldsymbol a}
	(\sum_{t=1}^T
	\boldsymbol p_t
	\tilde Y_t)/\sigma^2
\end{eqnarray*}
where
$\boldsymbol p_t = (\cos(\omega t), \sin(\omega t))^T$,
$\boldsymbol \Omega = \text{diag}(\sigma_{a_1}^2, \sigma_{a_2}^2)$,
$\tilde Y_t = Y_t - \mu - \psi \times ((-\log(\Phi(\beta_t)))^{-\xi} - 1)/\xi$.
To conduct an MCMC algorithm for this model, 
we could adapt the algorithm described in Section~\ref{sec-3.5},
adding an extra step to sample $\boldsymbol a$.

Sometimes, a dataset may show seasonality in the scale parameter,
such as in a weather dataset, the variation in the winter
may be larger than in the summer. 
Our model has the flexibility to deal with a dataset like this by
adding two similar sinusoids to $\psi$,
i.e. $\psi + a_3 \cos(\omega t) + a_4 \sin(\omega t)$.
It is possible to add sinusoids to the $\xi$ parameter
or adding more sinusoidal components to each parameter;
however, more parameters in the model will reduce the mixing rate
between chains in the MCMC algorithm;
as we will shown in the simulation study, 
the dGEV model already exhibit a large autocorrelation between
samples in the chain.
For simplicity,
we add seasonality only to the location parameter.
In the later context, 
we will refer this model to be the seasonal dGEV model;
however, it should be kept in mind that seasonality could be added to
any of the parameters.

\section{Illustrative simulation study}
\label{gev-sim}

In this section, we illustrate the dGEV model 
and seasonal dGEV model (\ref{equ-seasonal-mean}) using simulated data.
To generate datasets, 
we start by generating time-varying parameters $\beta_{1:T}$,
and then use $\beta_{1:T}$ to generate observations $Y_{1:T}$.

To be more specific, 
for the dGEV model, we generate 1000 observations.
The choice of the sample size is arbitrary;
however, more observations are rquired for a model with more parameters,
for example in the case where we add seasonality to
both the location and scale parameters.
In the simulations, we choose
$\phi = 0.8$,
$\mu = 0.5$, 
$\psi = 0.3$
$\xi = 0.05$,
and $\sigma = 0.1$.
We first simulate $\beta_1 \sim N(0, 1)$,
then $\beta_2, \dots, \beta_T$ can be generated according to the AR(1) process
in (\ref{equ-beta-ar1}).
Based on the generated values of $\beta_{1:T}$,
we then generate dataset $Y_{1:T}$ from (\ref{equ-observation}).
For the seasonal dGEV model~\ref{equ-seasonal-mean}, 
we choose
$a_1 = 1$,
$a_2 = 2$,
and suppose we have an annual dataset,
$f = 1 / 365.25$ and $\omega = 2 \pi / 365.25$.

After generating observations, 
we need to chose priors to conduct the Bayesian analysis.
The priors for those parameters are as follows:
$\mu~\sim~N(0, 2^2)$,
$\psi~\sim~\text{Gamma}(2, 2)$,
$\xi~\sim~N(0, 2^2)$,
$\phi~\sim~\text{uniform}(-1, 1)$,
and
$\sigma^2~\sim~\text{IG}(1, 0.01)$.
We choose the total iteration for MCMC to be 20,000,
when running the algorithm,
with the first 5,000 as burnin.
In the PGAS step, we need to choose number of particles;
here we choose the number to be 1000. 
\citet{lind14} shows that when this rate is chosen,
the PGAS algorithm has update frequency very close to 0.999,
the ideal rate.

\begin{figure}
\centering
\includegraphics[width=14cm]{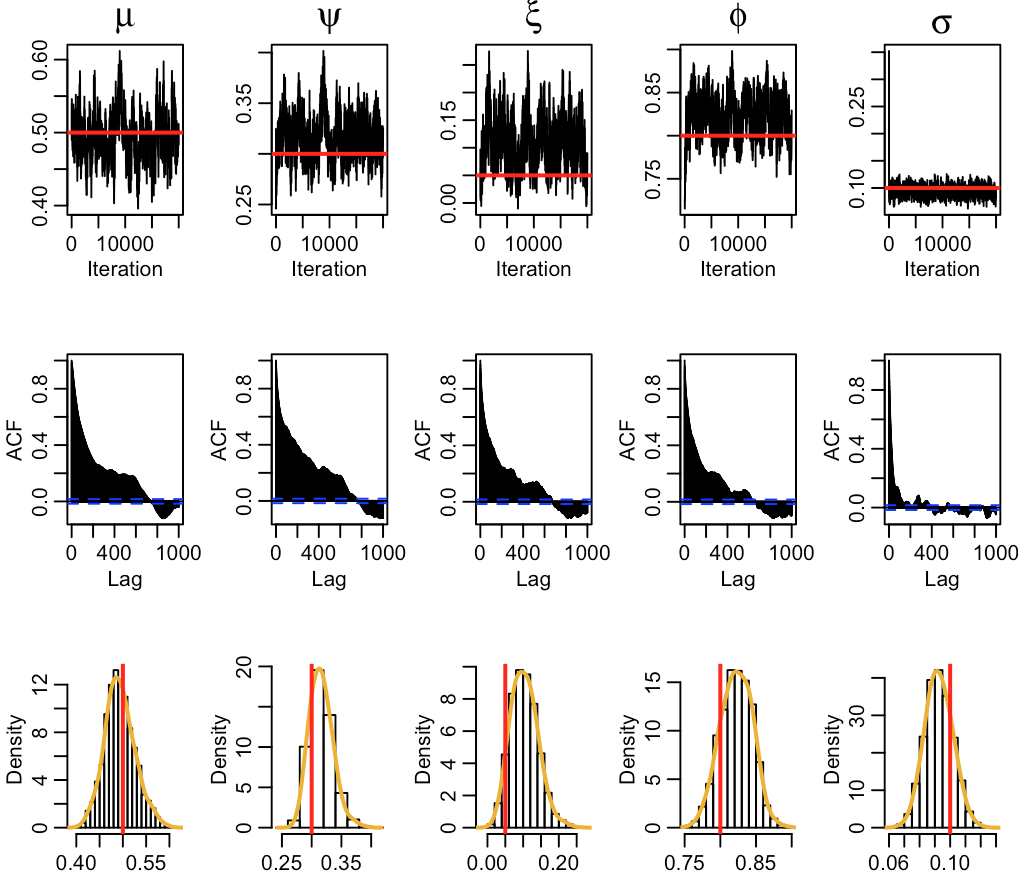}	
\caption{Plot of MCMC draws (1st row), autocorrelation function (ACF) (2nd row) 
				and histogram with density (3rd row)
				for $\mu$, $\psi$, $\xi$, $\phi$, $\sigma$.
				The red line indicates the true values.}
\label{fig-draws-dGEV}
\end{figure}

We then fit the model to the simulated datasets.
The first row in Figure~\ref{fig-draws-dGEV}
plots the MCMC draws for parameters except for 
$\beta$s for the dGEV model.
The draws converge quickly for each parameter;
the red lines that indicate the true values are all contained in the draws. 
Table~\ref{tab-sim-posterior-dGEV} gives summary statistics for the posterior
medians and
credible intervals. 
\begin{table}
\caption{Posterior medians and 95\% credible intervals for parameters 
				$\mu$, $\psi$, $\xi$, $\phi$, $\sigma$.}
\label{tab-sim-posterior-dGEV}
\centering
\begin{tabular}{ccccc}
\toprule
Parameter & True value  & Posterior median & 95\% Credible interval &
    Inefficiency factor \\
\midrule
$\mu$            & 0.5 & 0.492  & [0.4332, 0.5642]  & 222.26\\
$\psi$             & 0.3  & 0.315 & [0.2818, 0.3599]  & 233.84\\
$\xi$                & 0.05  & 0.101 & [0.035, 0.1843]  & 186.26\\
$\phi$             &  0.8 & 0.823 & [0.7781, 0.8636] &  173.69\\
$\sigma$        & 0.1 & 0.092 & [0.0753, 0.1109] & 61.61
\\
\bottomrule
\end{tabular}
\end{table}
The results show that the true values lie within the 95\% credible intervals,
and
the posterior medians are close to their true values, except for the parameter 
$\xi$. 
The issue with $\xi$, the shape parameter in the GEV distribution,
is that the proposed distribution is a less accurate approximation
to the true posterior distribution than in the case of
the location and scale parameters;
thus its the posterior median has larger bias compared with the
posterior medians of $\mu$ and $\psi$.

Table~\ref{tab-sim-posterior-dGEV} also provides the inefficiency factor for
each parameter;
this statistic gives a diagnostic about how well the chains in MCMC mixed.
It is calculated as $1 + 2 \sum_{s=1}^M (1-s/M) \rho_s$,
as given by \citet{chib01},
where $\rho_s$ is the estimated autocorrelation at lag $s$,
and $M$ is the batch size, which we take as 500. 
An inefficiency factor of $m$ means that the
effective number of draws is the
total number of iterations, after deleting burnin, divided by $m$.
The results in the table show that those inefficiency factors are large,
which corresponds to the slow decay in the ACF plots in Figure~\ref{fig-draws-dGEV}.

\begin{figure}
\centering
\includegraphics[width=13cm]{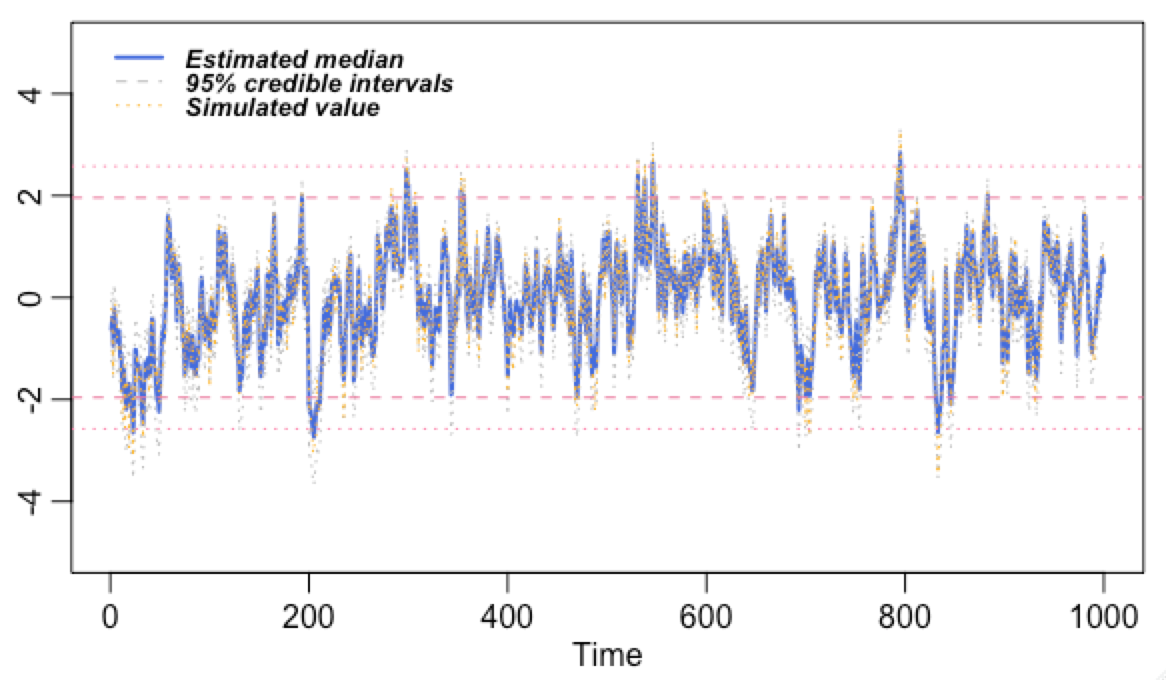}	
\caption{Plot of posterior median (blue line) and 95\% credible intervals (grey dash lines) 
				for $\beta_1, \dots, \beta_{1000}$;
				the yellow dashed lines indicates our simulated value for $\beta$s;
				the four dash red lines from up to down represent the 0.995, 0.975, 0.025, 
				0.005 quantiles of the standard normal distribution.}
\label{fig-beta-dGEV}
\end{figure}

Figure~\ref{fig-beta-dGEV} plots the posterior median and 95\% credible intervals 
for $\beta_1, \dots, \beta_{1000}$. 
Recall that the time-varying parameters are in a standard normal copula;
thus each $\beta_t$ has a marginal standard normal distribution.
Examining the posterior medians of $\beta$s shows that the
majority of them fall in the 95\% standard normal interval. 
The credible intervals are very small around each posterior median, 
as a result of the small value of measurement error $\epsilon_t$.

\begin{figure}
\centering
\includegraphics[width=14cm]{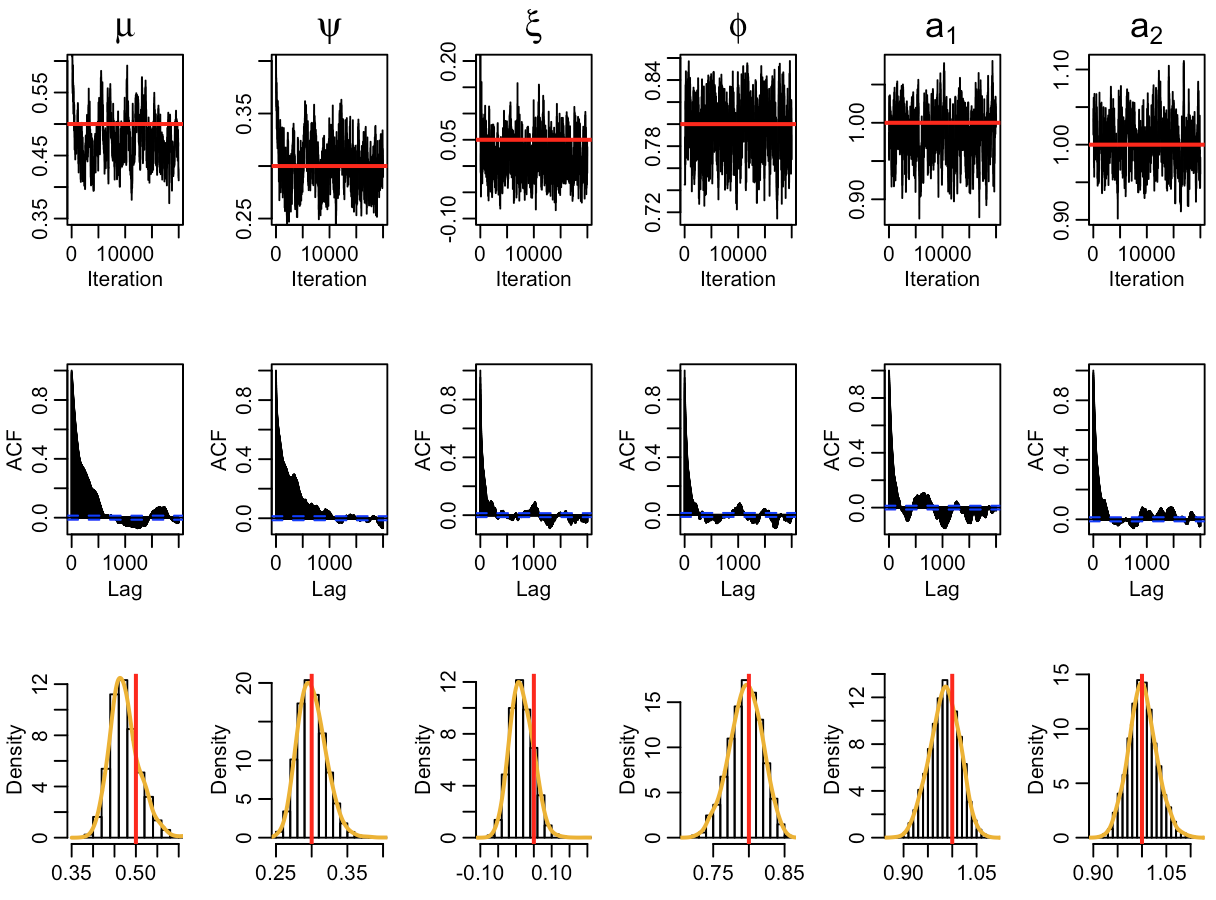}	
\caption{Plot of MCMC draws (1st row), autocorrelation function (ACF) (2nd row) 
				and histogram with density (3rd row)
				for $\mu$, $\psi$, $\xi$, $\phi$, $a_1$, $a_2$.
				The red line indicates the true values.}
\label{fig-draws-dGEV-season}
\end{figure}

\begin{figure}
\centering
\includegraphics[width=13cm]{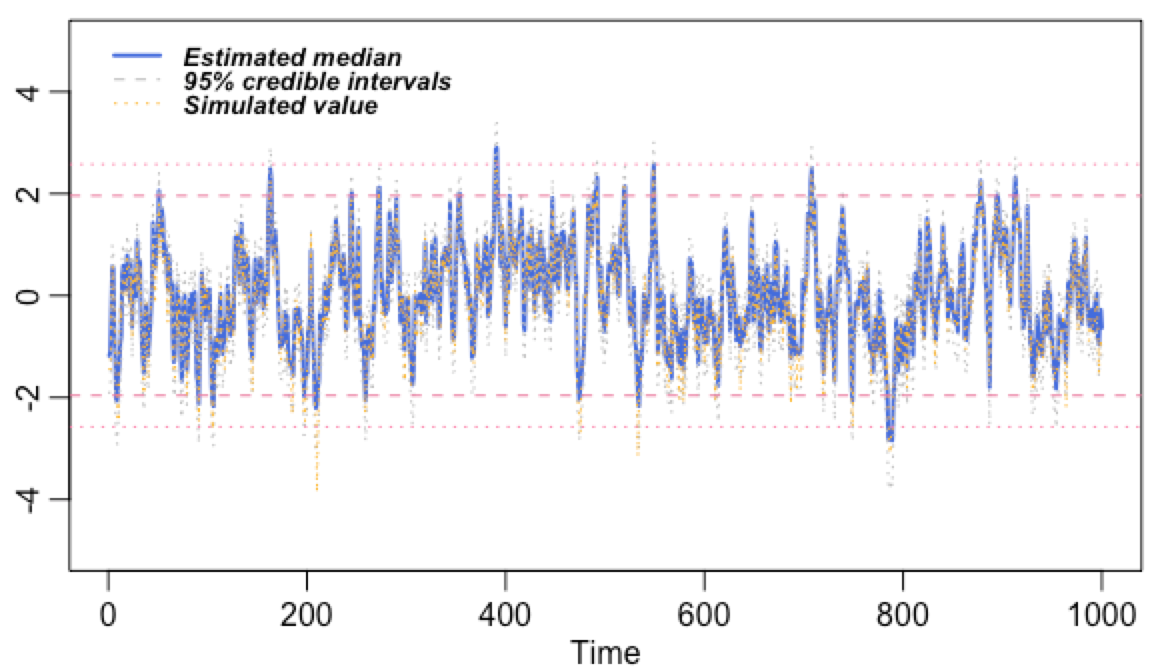}
\caption{Plot of posterior median (blue line) and 95\% credible intervals
    (grey dash lines) for $\beta_1, \dots, \beta_{1000}$;
    the yellow dashed lines indicates our simulated value for $\beta$s;
    the four dash red lines from up to down represent the 0.995, 0.975, 0.025, 
    .005 quantiles of the standard normal distribution.}
\label{fig-beta-dGEV-season}
\end{figure}

Figure~\ref{fig-draws-dGEV-season} shows the simulation results of 
selected parameter draws based on the seasonal dGEV model.
The estimated posterior median and 95\% credible intervals of
those parameters are summarized in Table~\ref{tab-sim-seasonal-dGEV}.
\begin{table}
\caption{Posterior median and 95\% credit intervals for parameters 
				$\mu$, $\psi$, $\xi$, $\phi$, $a_1$, $a_2$, $\sigma$.}
\label{tab-sim-seasonal-dGEV}
\centering
\begin{tabular}{ccccc}
\toprule
Parameter & True value  & Posterior median & 95\% Credible interval &
    Inefficiency factor \\
\midrule
$\mu$            & 0.5 & 0.472  & [0.4172, 0.5485]  & 219.05\\
$\psi$             & 0.3  & 0.300 & [0.2681, 0.3396]  & 214.55\\
$\xi$                & 0.05  & 0.014 & [$-0.0432$, 0.08136]  & 92.22\\
$\phi$             &  0.8 & 0.797 & [0.7486, 0.8380] &  103.36\\
$a_1$             & 0.1 & 0.984 & [0.9212, 1.0409] & 115.30\\
$a_2$             & 0.1 & 1.000 & [0.9453, 1.0641]  & 129.67\\
$\sigma$        & 0.1 & 0.092 & [0.0775, 0.1064] & 101.07
\\
\bottomrule
\end{tabular}
\end{table}
The 95\% credible intervals all covered the true value,
and posterior medians are close to their true value.
Similar to the non-seasonal model,
$\psi$ has larger bias compared with $\mu$ and $\psi$.
We also found the inefficiency factor for parameters in the seasonal dGEV model
to be on average higher than the non-seasonal model;
this is because having more parameters in the model increases the
correlations between chains.
Figure~\ref{fig-beta-dGEV-season} gives a plot of posterior medians
and their credible intervals for $\beta_1, \dots, \beta_{100}$. 
The figures show that the majority of draws fall in the 95\% region of the
standard normal distribution.
Also we plot the simulated true data of $\beta_1, \dots, \beta_{1000}$ in
Figure~\ref{fig-beta-dGEV};
all the true values are covered by their estimated 95\% credible intervals.

\FloatBarrier

\section{Real data study}
\label{gev-rea}

We will conduct a study of two real datasets in this section.
The first dataset is an annal maximum water flow dataset and 
the second is a
minimum log-return dataset for the S\&P 500 stock index. 
We fit the dGEV model to both datasets,
and we fit the seasonal dGEV model to the S\&P 500.

\subsection{Water flow data}

\begin{figure}
\centering		
\includegraphics[width=16cm]{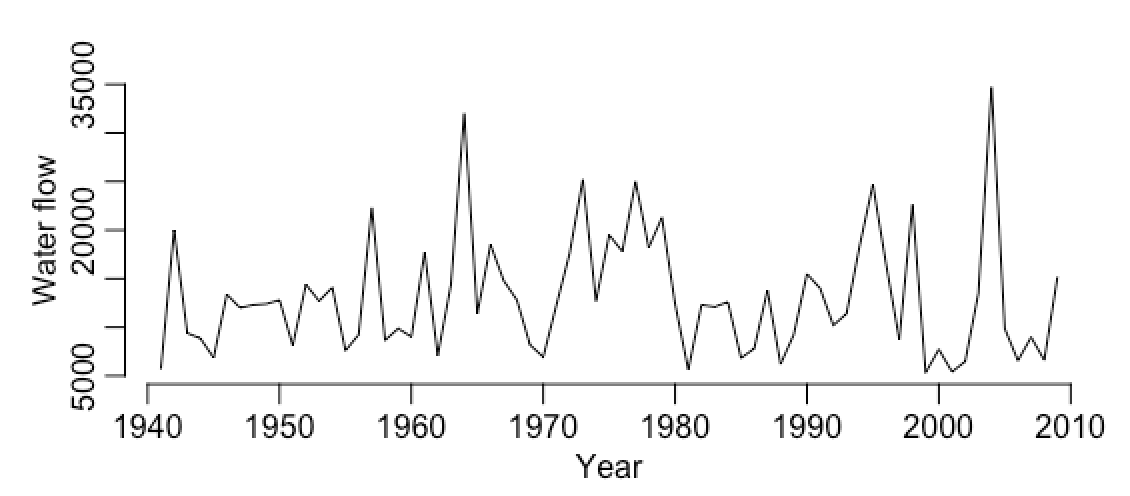}	
\caption{Annual maximum water flow of French Broad River at Asheville, North Carolina.}
\label{fig-river-data}	
\end{figure}

A water flow dataset is collected from French Broad River at Asheville in 
North Carolina.
The datasets contains annual maximum water flow level from 1941 to 2009.
A plot of this dataset is shown in Figure \ref{fig-river-data},
the plot shows there are two spikes in the year
1964 and 2004 which may consider to be unusual years. 
Our goal is to investigate how extreme those two values could be by fitting
into the dGEV model.

\begin{figure}
\centering
\includegraphics[width=12cm]{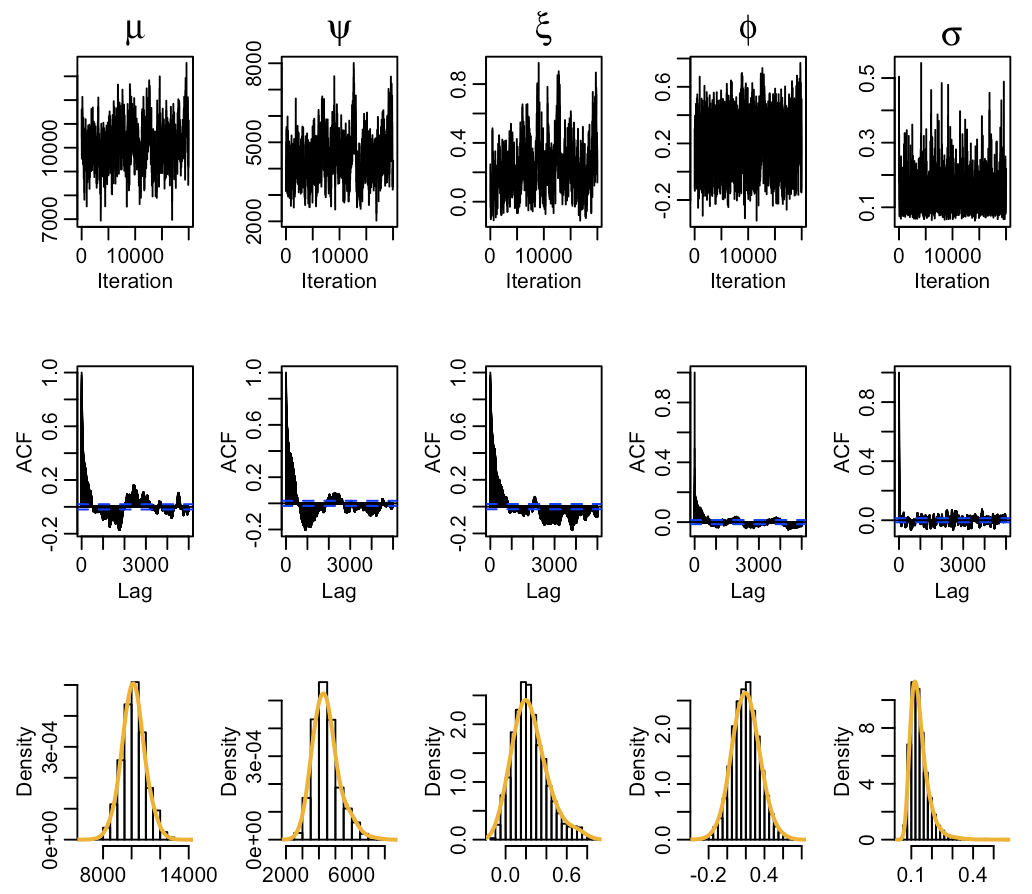}	
\caption{Plot of MCMC draws (1st row), autocorrelation functions (ACF) (2nd row) 
				and histograms along with densities (yellow lines) (3rd row)
				for parameters $\mu$, $\psi$, $\xi$, $\phi$, $\sigma^2$ in dGEV model
				of the water flow dataset.}
\label{fig-draws-river}			
\end{figure}

We run the chain for 20,000 draws and treat the first 5,000 as burnin,
the number of particles is chosen to be 1000. 
Before running the analysis, we standardize the dataset to avoid computation problem.
The plot of draws and summary statistics is presented in Figure
\ref{fig-draws-river} and Table \ref{tab-6.1}. 
The ACF plots and inefficient factors for each parameters shows the chain mixed well.

\begin{figure}
\centering
\includegraphics[width=14cm]{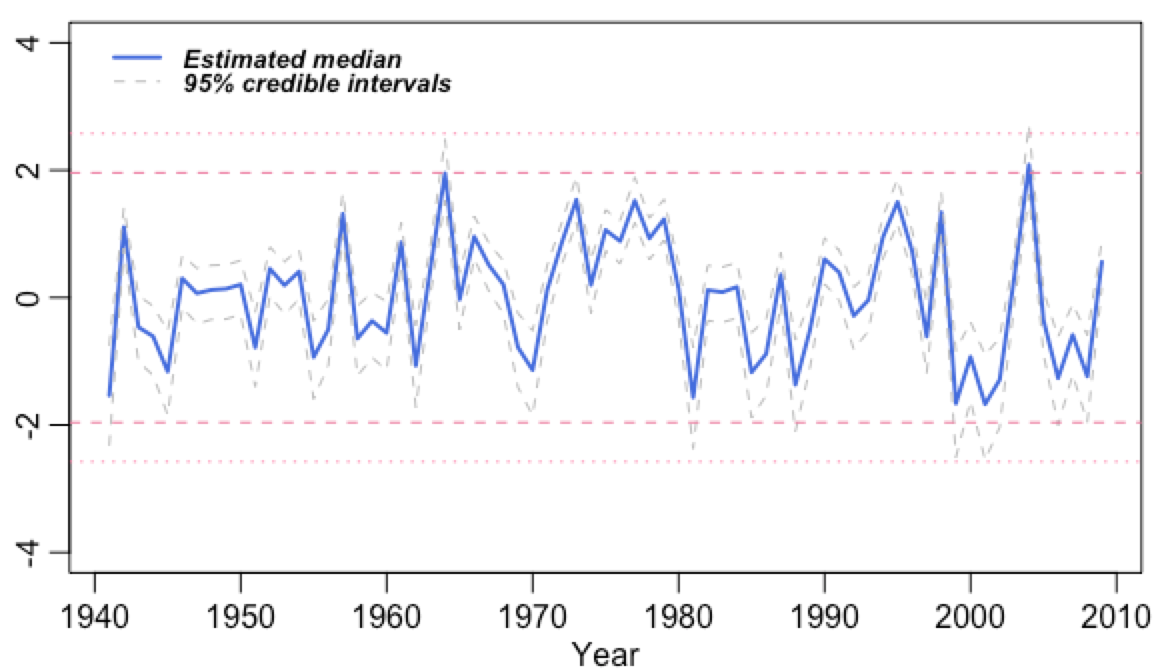}	
\caption{Plot of posterior medians (blue line) and their 95\% credible intervals (grey dash lines) 
				for $\beta_t$s by fitting dGEV into water flow dataset. 
				The four dash red lines from up to down represent the 0.995, 0.975, 0.025, 0.005 
				quantiles of the standard normal distribution.}
\label{fig-beta-river}			
\end{figure}

\begin{table}
\caption{Posterior medians, 95\% credible intervals and inefficient factors for parameters 
				$\mu$, $\psi$, $\xi$, $\phi$, $\sigma$.}
\label{tab-6.1}
\centering
\begin{tabular}{ccccc}
\toprule
Parameter  & Posterior median & 95\% Credible interval & Inefficiency factor \\
\midrule
$\mu$            & 10144  & [8566.1, 118645.0]  & 170.58\\
$\psi$             & 4351 & [3148.0, 6303.0]  & 223.22\\
$\xi$                & 0.22 & [$-0.031$, 0.664]  & 223.22\\
$\phi$             & 0.21 & [$-0.075$, 0.523] &  74.16\\
$\sigma$        & 10.27 & [6.468, 19.664] & 34.59 \\
\bottomrule
\end{tabular}
\end{table}

From Table \ref{tab-6.1},
the time varying parameter $\phi$ has posterior median 0.021,
the 95\% credible intervals cover with 0, 
the result suggests the dependency between extreme values are low or
possibly does not exist.

In Figure \ref{fig-beta-river}, we plot the posterior median for $\beta_t$s.
Since each $\beta_t$s has marginal distribution from standard normal,
it is convenient to compare these values with standard normal quantiles.
If the values is larger than $2.326$ in absolute value,
then it is beyond 99\% quantiles of normal distribution,
which may seems unusual. 
From the plot, both of the two events in 1964 and 2004 we are interested 
in lies on the boundary of 95\% quantile,
and their credible intervals are below 99\% quantile,
which may shows the two events are not extreme as it looks like from 
the data plot.

\subsection{S\&P 500 datasets}

\begin{figure}
\centering		
\includegraphics[width=14cm]{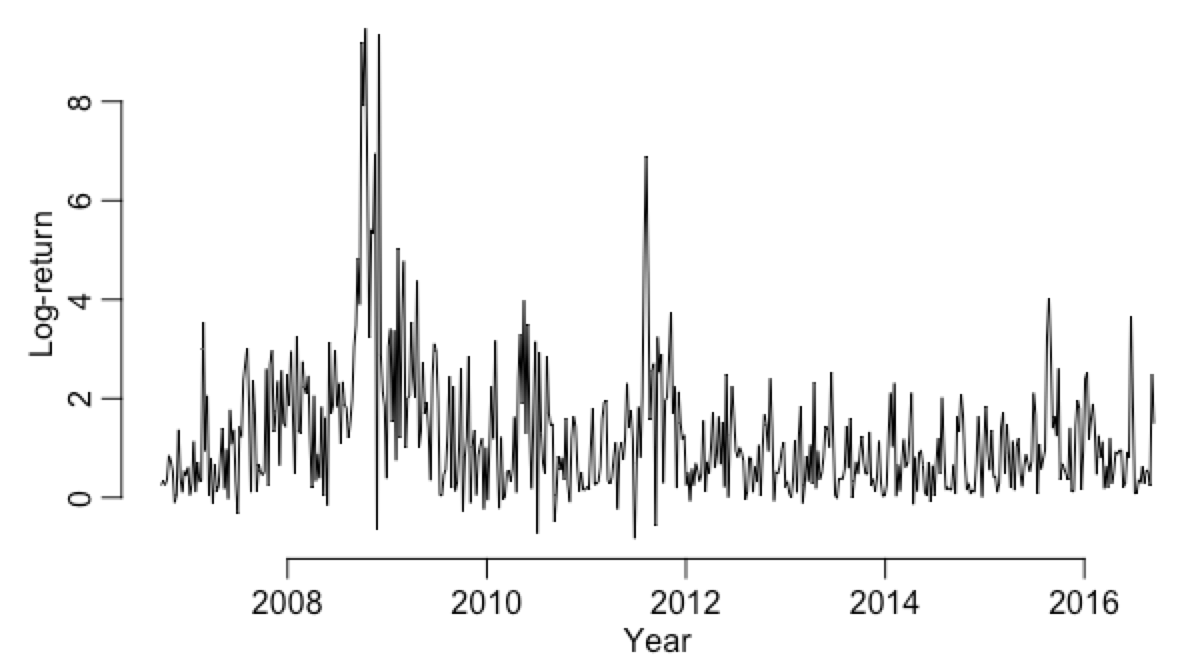}	
\caption{Weekly minimum S\&P 500 log-return dataset, adjust the dataset to be $-1$.}
\label{fig-sp500}	
\end{figure}

\begin{figure}
\centering	
	\includegraphics[width=11cm]{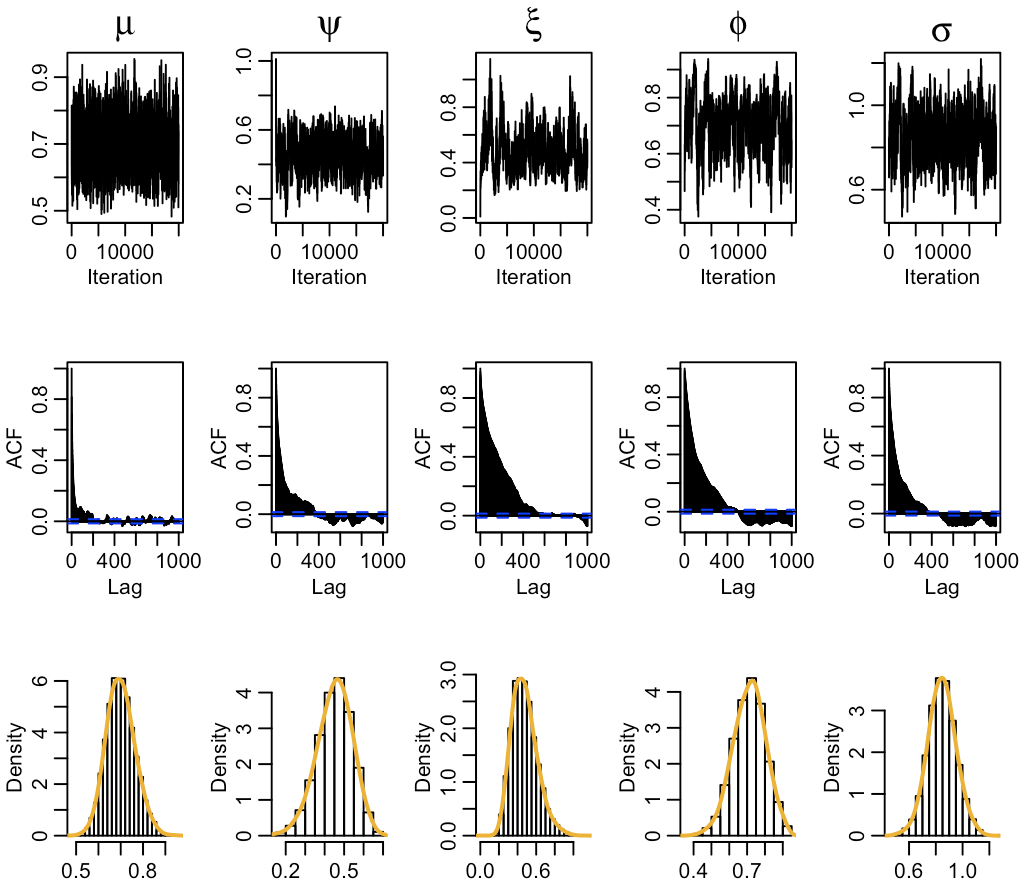}
	\caption{Plot of MCMC draws (1st row), autocorrelation functions (ACF) (2nd row) 
				and histograms along with densities (yellow line) (3rd row)
				for parameters $\mu$, $\psi$, $\xi$, $\phi$, $\sigma^2$ in the dGEV model 
				by fitting the S\&P 500 dataset.}
\label{fig-draws-sp500}		
\end{figure}

\begin{figure}
\centering	
	\includegraphics[width=12cm]{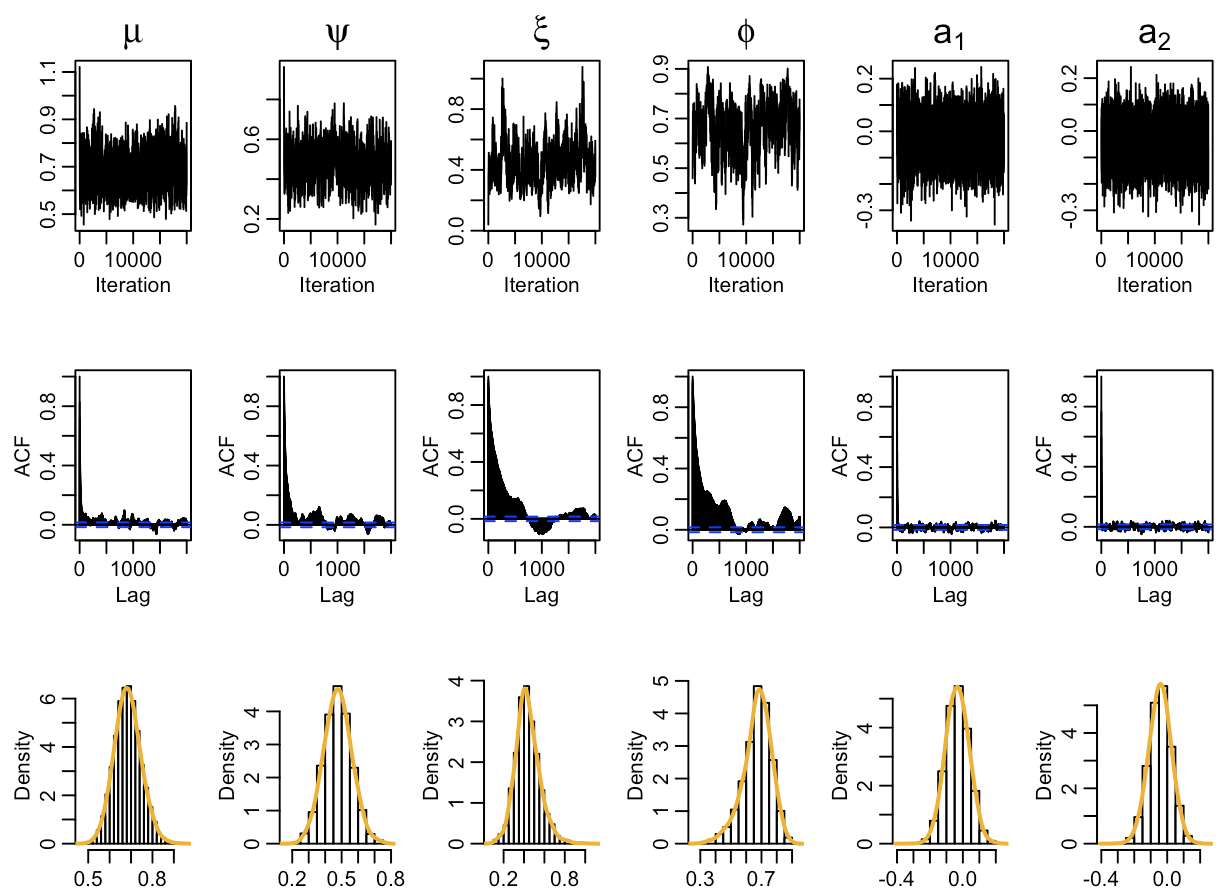}
	\caption{Plot of MCMC draws (1st row), autocorrelation function (ACF) (2nd row) 
				and histograms along with densities (yellow line) (3rd row)
				of parameters $\mu$, $\psi$, $\xi$, $\phi$, $a_1$, $a_2$ in a seasonal dGEV model
				by fitting the S\&P 500 dataset.}
\label{fig-draws-sp500-season}		
\end{figure}

Another dataset is from S\&P 500 minimum weekly log-return,
the data is collected from
September, 25, 2006 to September, 12, 2016. 
A plot of this data is shown in Figure \ref{fig-sp500}.
The data is plotted in a $-1$ scale,
so the largest value in the plots stands for the minimum return.
Our interests is focus on finding the most unusual happened events in the dataset,
especially at the period during 2008 to 2010.
Sometimes a time series shows seasonality,
although the seasonality is not so obviously to appear in this model,
we are interested to fit model to find out if seasonal effects exists.
We fit the dataset in both of the dGEV model and the seasonal dGEV model,
the plot of draws are shown in Figure \ref{fig-draws-sp500}
and Figure \ref{fig-draws-sp500-season}.
The posterior medians with their 95\% credible intervals for parameters
in the model are summarized
in Table \ref{tab-6.2}. 
The inefficient factors for those parameters are also provided.
In the MCMC algorithm, we run 20,000 iterations and treat the 5000 draws as burnin;
we choose the number of particles to be 1000 for both of the two models.
For the seasonal dGEV model, we choose $f = 7/365.25$.

From the result shown in Table \ref{tab-6.2},
the estimation for location, scale and shape parameters for both dGEV 
and seasonal dGEV model are close,
the inefficient factors for seasonal dGEV model are larger than the dGEV model.
In the output of the estimation of seasonality parameters $a_1$, $a_2$,
there value are very close to $0$, 
and their 95\% credible intervals contains 0. 
This suggests seasonality may not exists in this model. 

\begin{table}
\caption{Posterior medians with their 95\% credible intervals
              and inefficient factor for parameters 
				$\mu$, $\psi$, $\xi$, $\phi$, $\sigma$ in the dGEV model
				and 
				parameters $\mu$, $\psi$, $\xi$, $\phi$, $\sigma$, $a_1$,  $a_2$
				in the seasonal dGEV model.}
\label{tab-6.2}
\centering
\begin{tabular}{cccc}
\multicolumn{4}{c}{{\bf Parameter estimates for dGEV model}} \\
\toprule
Parameter  & Posterior median & 95\% Credible interval & Inefficient factor \\
\midrule
$\mu$            & 0.696  & [0.5847, 0.8307]  & 37.83\\
$\psi$             & 0.348 & [0.2171, 0.4645]  & 80.29\\
$\xi$                & 0.449 & [0.2638, 0.7264]  & 179.25\\
$\phi$             & 0.713 & [0.5347, 0.8555] &  145.46\\
$\sigma$        & 0.967 & [0.7425, 1.1817] & 96.37 \\
\bottomrule
\end{tabular}
    \\[2ex]
\begin{tabular}{cccc}
\multicolumn{4}{c}{{\bf Parameter estimates for seasonal dGEV model}} \\
\toprule
Parameter  & Posterior median & 95\% Credible interval & Inefficient factor \\
\midrule
$\mu$            & 0.683  & [0.5691, 0.8101]  & 39.16\\
$\psi$             & 0.476 & [0.3132, 0.6440]  & 91.94\\
$\xi$                & 0.425 & [0.2294, 0.6912]  & 220.46\\
$\phi$             & 0.680 & [0.4547, 0.8277] &  201.65\\
$a_1$             & $-0.034$ & [$-0.1712$, 0.1027]    & 7.75\\
$a_2$             & $-0.042$ & [$-0.1794$, 0.0890] & 7.76\\
$\sigma$        & 0.531 & [0.3729, 0.6480] & 154.70\\
\bottomrule
\end{tabular}
\end{table}

\begin{figure}
\centering
   \begin{subfigure}[a]{0.8\textwidth}
   \includegraphics[width=1\linewidth]{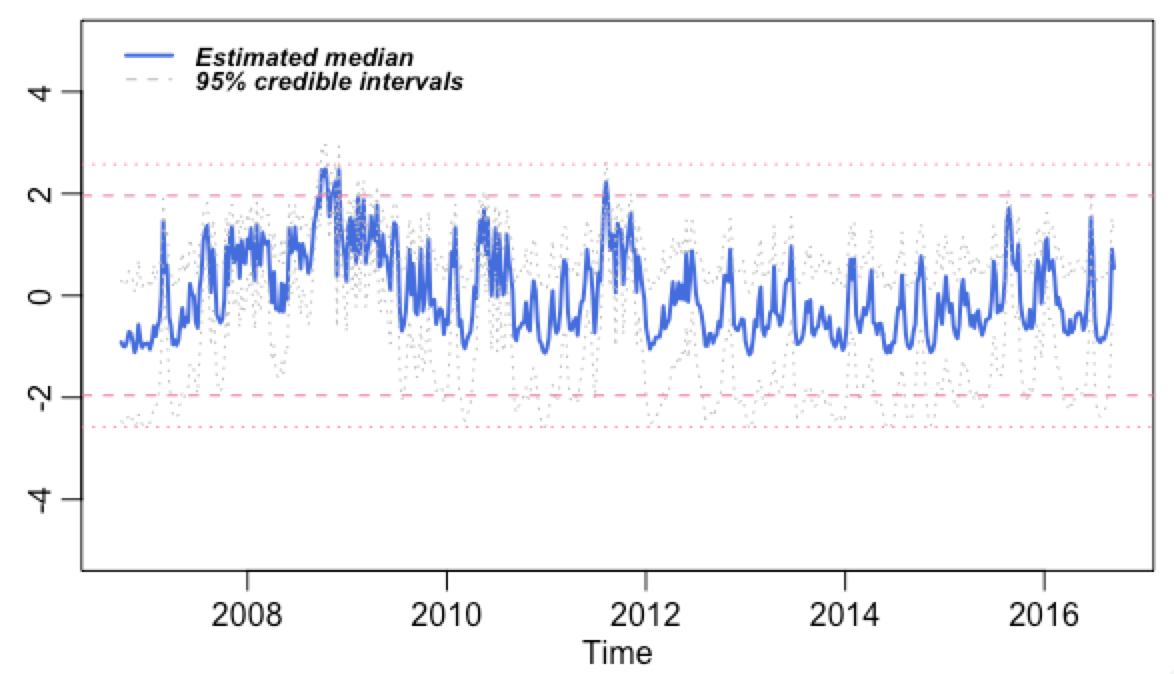}
   \caption{}
   \label{fig-beta-sp500} 
\end{subfigure}

\begin{subfigure}[b]{0.8\textwidth}
   \includegraphics[width=1\linewidth]{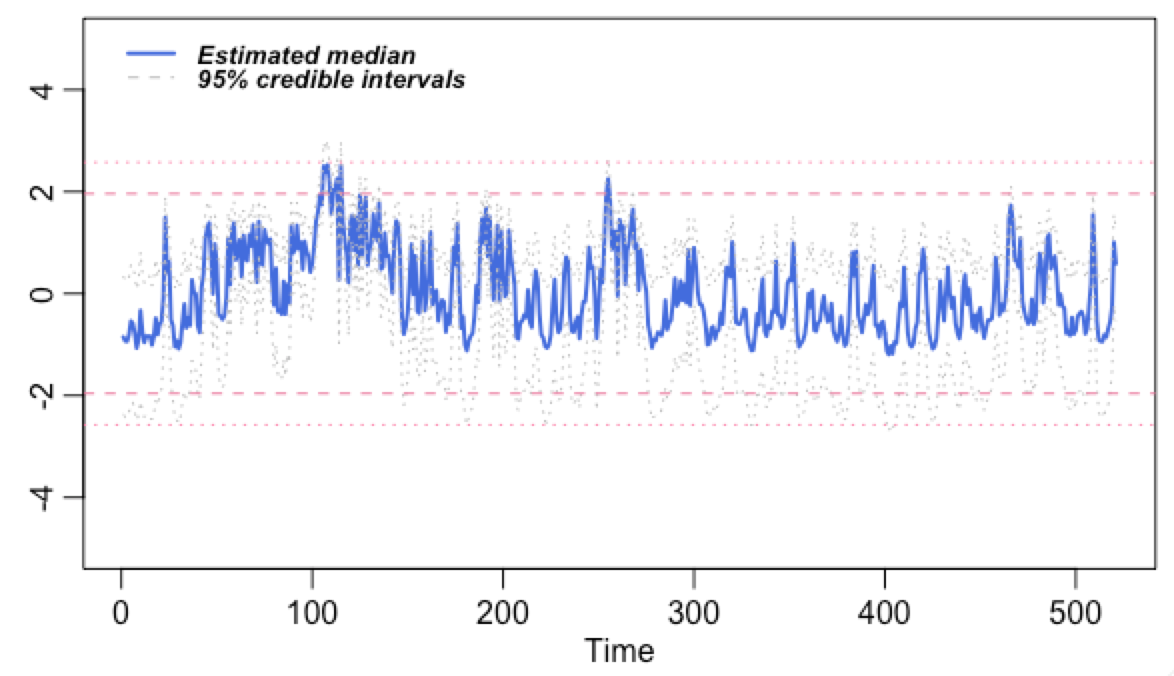}
   \caption{}
   \label{fig-beta-sp500-season}
\end{subfigure}

\caption{
(a) Plot of posterior medians (blue line) and their 95\% credible intervals 
(grey dash lines) for $\beta_t$s by fitting the dGEV model into S\&P 500 dataset. 
(b) Plot of posterior medians (blue line) and their 95\% credible intervals 
(grey dash lines) for $\beta_t$s by fitting the seasonal dGEV model into S\&P 500 dataset. 
The four dash red lines from up to down represent the 0.995, 0.975, 0.025, 0.005 quantiles of the standard normal distribution. dataset.}
\label{fig-beta-sp500-both}
\end{figure}

Figure \ref{fig-beta-sp500-both} gives plot of estimated
$\beta_t$s posterior medians and their 95\% credible intervals for
fitting both models. 
The two models gives a very similar estimates for those $\beta_t$s. 
Both plots indicates the period during 2008 crisis are much more 
unusual than other period. 
The highest point has its posterior median close to 99\% quantile of 
standard normal, 
and its credible intervals bounds contains the area which exceed the 99\%
quantile. 

\FloatBarrier

\section{Conclusion and discussion}
\label{gev-con}

In this paper, we propose a novel dependent GEV model. 
The model can be expressed as nonlinear Gaussian state space model.
Due to the observation equation is high nonlinear, 
we use PGAS to sample time-varying 
parameters. 
and incorporate it into an MCMC algorithm.
The simulation study based on the MCMC algorithm shows the 
sampled parameter distribution could cover the true value
and the posterior median is close to the truth.
The shape parameter $\xi$ is turned to be harder to estimate
compare to the location and scale parameters.
We also show our dGEV model can easily to incorporate seasonal 
components.
Seasonality can be added to both the location, scale and shape parameters.

We did two case studies for the model: one is the annual maximum
water flow dataset and another is the weekly minimum return of S\&P 500. 
The estimated dependent parameter in the water flow dataset contains value 0
which suggests the correlation between extreme values are low.
The two unusual events in 1964 and 2004 does not turn out to be very unusual 
after plotting the corresponding estimated value of $\beta$.
In the S\&P 500 dataset, the correlation is very high.
By fitting this dataset into the seasonal dGEV model which suggests there
does not exist any seasonality effect. 
The draws of $\beta_t$s shows extreme values during the 2008 economics crisis 
is very unusual compares to other values. 

Our model turns out to require large computation time when the sample
size is larger. 
However, the sample size usually becomes 10,000 when analyzing 
extremely daily values for temperature. 
In order to analyze those datasets, 
a improvement of current algorithm is need. 
On the other side, 
for datasets contains seasonality on the scale or shape parameters,
the number of parameters will increase and cause the inefficient factors
become larger. 
Thus a method to redesign the MCMC algorithm is needed as well. 
Both of these issues need to be address in the future study but
is beyond the scope of this paper.

%
%
%
%
%
%

\bigskip

\bibliographystyle{apalike}

\bibliography{citation}

\end{document}